\documentclass[journal]{IEEEtran}
\usepackage{cite}

\ifCLASSINFOpdf
\else
\fi
\usepackage{amsmath}

\input{MySettings.tex}

\begin{document}
%
\title{A Corrugated Planar-Goubau-Line Termination for Terahertz Waves}
%
%
%

\newcommand{\orcidauthorA}{0000-0002-9821-3199} 
\newcommand{\orcidauthorB}{0000-0002-5861-6655} 
\newcommand{\orcidauthorC}{0000-0001-9519-3527} 
\newcommand{\orcidauthorD}{0000-0002-8204-7894} 

\author{Juan~Cabello-S\'{a}nchez,~\IEEEmembership{Member,~IEEE,}
Vladimir~Drakinskiy,
Jan~Stake,~\IEEEmembership{Senior~Member,~IEEE,}
Helena~Rodilla,~\IEEEmembership{Senior~Member,~IEEE}
\thanks{Manuscript received 18\textsuperscript{th} December, 2022; revised 22\textsuperscript{th} January, 2023; accepted 23\textsuperscript{rd} January, 2023. This work was supported by the Swedish Research Council (VR) under grant 2020-05087.}
\thanks{Juan Cabello-S\'{a}nchez, Vladimir Drakinskiy, Jan Stake and Helena Rodilla are with the Terahertz and Millimetre Wave Laboratory, Chalmers University of Technology, SE-412 96 Gothenburg, Sweden. (e-mail: \mbox{cabellosanchez.juan@gmail.com}; \mbox{vladimir.drakinskiy@chalmers.se}; \mbox{jan.stake@chalmers.se}; \mbox{rodilla@chalmers.se})}
\thanks{Color versions of one or more of the figures in this article are available online at http://ieeexplore.ieee.org.}
\thanks{Digital Object Identifier \href{http://dx.doi.org/}{XXXXXXXXXX} }}%
%
%

\markboth{IEEE Microwave and Wireless Technology Letters,~Vol.~XX No.~XX, XX~2023}%
{Cabello-S\'{a}nchez \MakeLowercase{\textit{et al.}}: Capacitively-coupled open-stub resonators for terahertz planar Goubau line filters}
%



\maketitle

\begin{abstract}
The planar Goubau line is a promising low-loss metal waveguide for terahertz applications. To enable advanced circuits and multi-port measurements based on planar Goubau lines, there is a strong need for broadband impedance-matched loads, which can be used to absorb the energy and minimize standing waves in a system. In this work, we propose a termination for planar Goubau lines based on an exponentially-tapered corrugated line, gradually increasing conductor losses while maintaining small reflections. The corrugation density is high enough to increase conductor losses without requiring an auxiliary low-conductivity material. A 400-µm long planar Goubau line load was fabricated on a 10-$\mu$m thick silicon substrate suspended in the air. Simulations of the load show excellent agreement with calibrated reflection measurements in the frequency range 0.5 THz – 1.1 THz. Above the cut-off frequency of around 580 GHz, the measured reflections are less than -19\,dB, below the noise floor of the characterization setup.

\end{abstract}


%
\IEEEpeerreviewmaketitle


\section{Introduction}

\IEEEPARstart{E}{fficient} and broadband planar waveguides are essential for integrated photonic and electronic systems \cite{Sengupta2018}.
However, traditional planar multi-conductor lines have a relatively high loss at terahertz frequencies.
One alternative is to use a single-conductor planar waveguide, the planar Goubau line (PGL).
The PGL consists of a single conducting strip deposited on a dielectric substrate\cite{Akalin2006}.
Similar to other single-wire waveguides like the Sommerfeld's wire \cite{Sommerfeld1899} and the Goubau line \cite{Goubau1950}, the PGL propagates a quasi-TM (hybrid EH) surface wave with a field that decays away from the surface of the metal strip.
The PGL has a low attenuation constant compared to other metal waveguides when designed on an electrically-thin substrate \cite{CabelloSanchez2018}, making it an efficient alternative for guiding terahertz waves.
Although several components have been proposed for PGL, the performance needs to be improved for using single-wire interconnects in applications.

\begin{figure}[t]
\includegraphics[width=0.9\linewidth]{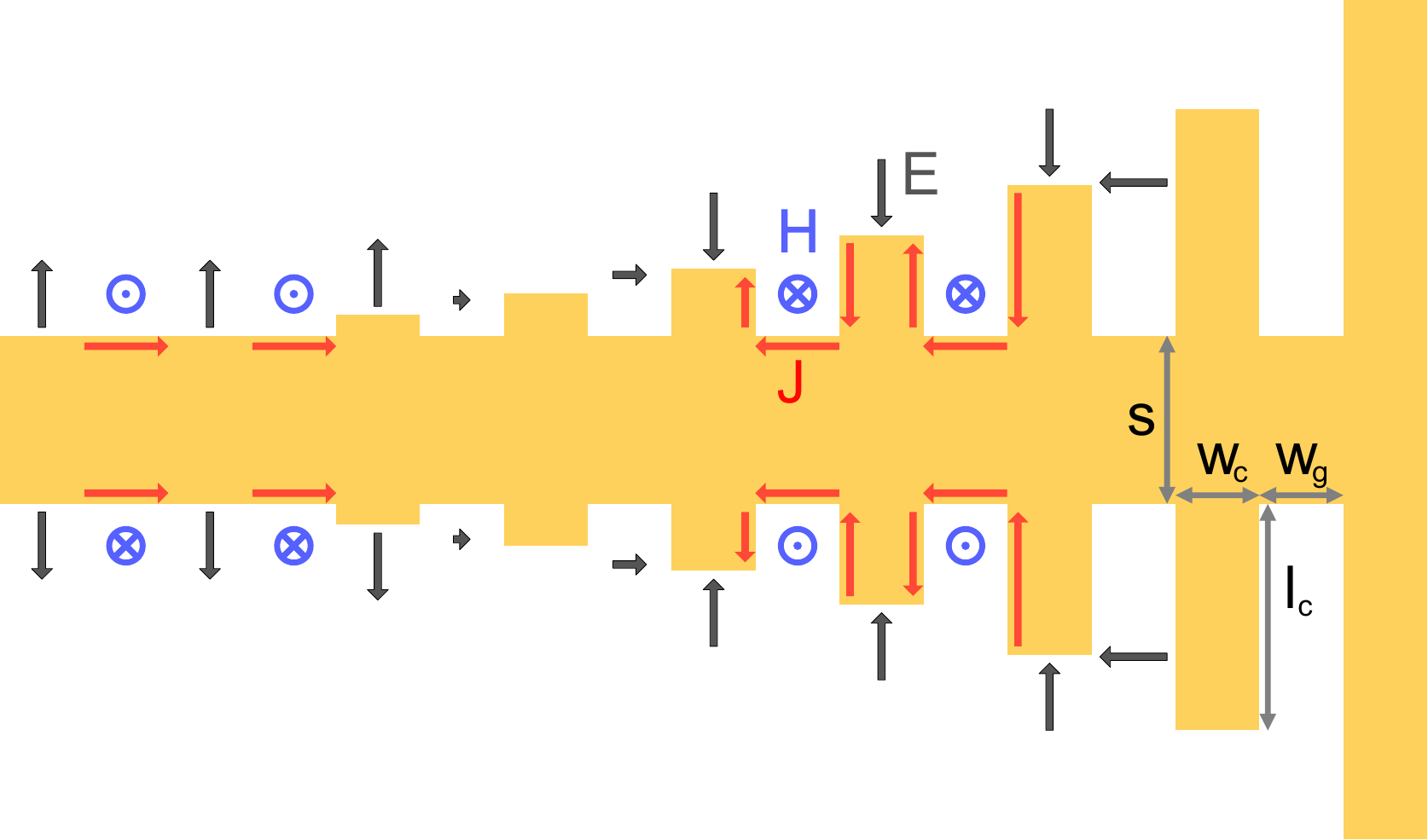}
\caption{\label{fig:Currents}Illustration of the tapered corrugated planar-Goubau-line load. The figure shows the orientation of the electric (black) and magnetic (blue) fields, and current densities (red). The fabricated load has $s=\SI{10}{\um}$, $w_c=\SI{1}{\um}$, $w_g=\SI{1}{\um}$ and $l_c$ going from \SI{1}{\um} to \SI{50}{\um} exponentially.}
\end{figure}

In recent years, some circuit elements---such as antennas \cite{Sanchez-Escuderos2013}, resonators \cite{Horestani2013a}, and filters \cite{CabelloSanchez2022}---have been proposed for PGL, including calibration standards \cite{Cabello-Sanchez2018}.
A vital circuit element is an impedance-matched load, which allows terminating ports with low reflections, enabling components and multi-port scattering-parameters measurements.
Matched loads can be done using lumped resistive elements at microwave frequencies. Still, their fabrication becomes increasingly difficult at higher frequencies, requiring other methods at THz frequencies, such as distributed elements along the waveguide. Some matched-load designs have been proposed for PGL at frequencies between 50\,GHz and 500\,GHz. Xu et al.\cite{xu2011} suggested a PGL load was made by terminating a line with a PGL-CPW transition with a short-circuited titanium-film resistor. Treizebre et al. \cite{Treizebre2012} proposed using a tapered nickel-chromium high-resistance sheet below the PGL, and Le Zang et al. \cite{LeZhang2022a} proposed using a low-conductivity nickel PGL with hook-shaped corrugations on one side of the line.

All these proposed loads partly rely on the low conductivity of a second thin-film material to increase the ohmic losses, which adds complexity to the fabrication process.
In this paper, we propose and demonstrate an impedance-matched load based on a corrugated PGL \cite{Laurette2012}, which has been smoothly-tapered to minimize reflections and has a high corrugation density to increase ohmic losses, see Fig.\,\ref{fig:Currents}.



\section{Method} \label{sec:method}


\begin{figure}[t]
    \centering
    \includegraphics[width=0.99\linewidth]{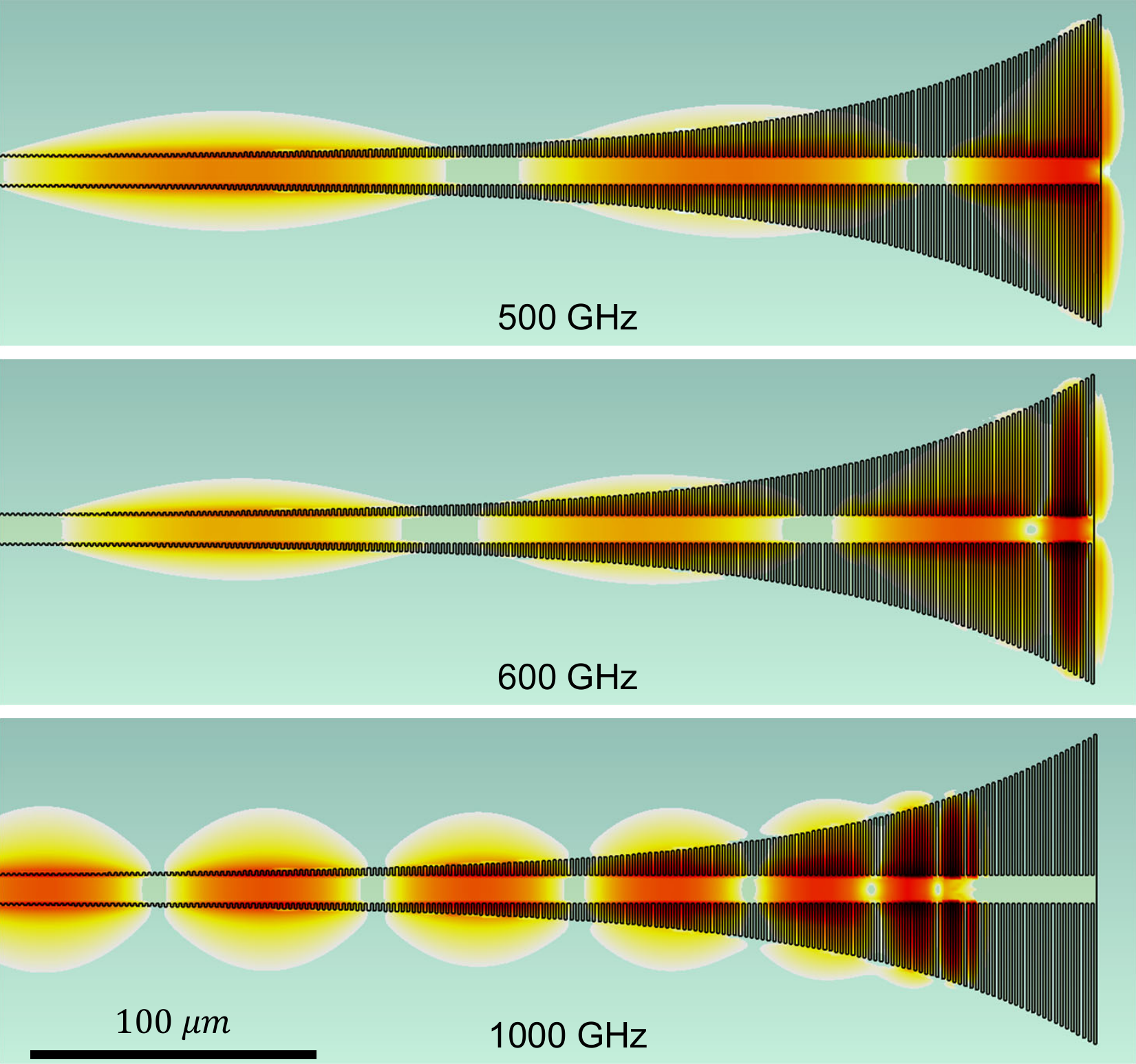}
    \caption{Simulated magnetic-field intensity of the PGL load. The field is shown at 500\,GHz (below the cut-off frequency), 600\,GHz (above the cut-off) and 1000\,GHz (well above the cut-off). The attenuation of waves increases dramatically when the corrugation length is close to $l_c=\lambda_0/(4n_c)$, where $n_c$ is the effective refractive index of the corrugations.}
    \label{fig:load_Hfields_3freqs}
\end{figure}

Corrugated waveguides can support TM surface waves if the corrugation length, $l_c$, is less than a quarter-wavelength, $\lambda_c/4$, of the wave in the corrugation \cite{Rotmant1950}.
When the corrugations are almost a quarter-wavelength long, the phase constant, the field confinement, and the conductor losses dramatically increase \cite{Rotmant1950,Elliott1954,ChapinCutler1994}. The conductor losses result from the surface currents induced by the difference of the magnetic field in the conductor's boundary, as stated by Ampère's law.
The power loss in the conductor can be calculated as \cite{Collin2001_emtheory}:
\begin{equation} \label{eq:induction}
    P_c = \int_S\frac{1}{2}\text{Re} \{ Z_m \} \lvert\Vec{H_t}\rvert^2 dS
\end{equation}
where  $H_t$ is the magnetic field tangent to the conductor, $Z_m=\sqrt{j\omega \mu/\sigma}$ is the intrinsic impedance of the metal, $j$ is the imaginary unit, $\omega$ is the angular frequency, and $\mu$ and $\sigma$ are the metal's permeability and conductivity, respectively.
Conductor losses can be maximized at a given frequency by increasing the metal's  $\mu/\sigma$ ratio or by increasing the surface of the metal exposed to the tangential magnetic field.
Thus, by using higher corrugation length, $l_c$, and lower corrugation width, $w_c$, and gap, $w_g$, the ohmic losses are increased dramatically without needing a second material with a lower $\mu/\sigma$ ratio, simplifying the fabrication process.
A sudden increase of losses in an electrically-short distance results in an input impedance close to a short circuit, increasing reflections.
Therefore, $l_c$ needs to increase gradually to minimize reflected power losses.
Moreover, a gradual increase in $l_c$ will increase the absorbing bandwidth of the corrugated PGL.
Reflections are also lowered if the magnetic-plane symmetry of the PGL's propagating mode is respected by corrugating both sides of the PGL. Finally, the proposed load's minimum and maximum frequency of operation can be tuned by choosing the longest and shortest $l_c$, respectively, where the $w_c$ and $w_g$ must be as narrow as possible to increase losses. In this work, the load design exhibits a corrugation length, $l_c$, that increases exponentially from \SI{1}{\um} to \SI{50}{\um} long, over a length of \SI{400}{\um}, whereas the corrugation width, $w_c$, and gap, $w_g$, are both \SI{1}{\um}, two orders of magnitude smaller than the wavelength. This results in a lower cut-off frequency of approximately \SI{580}{GHz} when $l_c$=$\lambda_0/(4n_c)=\SI{50}{\um}$, where $n_c=2.4$ is the effective refractive index of the waves propagating in the corrugations, approximated as a coplanar stripline odd mode. \textcolor{black}{To avoid substrate modes and radiation loss \cite{Riaziat1990a},} the design was implemented on a 10-$\mu$m thick silicon substrate suspended in the air.

Full electromagnetic simulations were carried out using a time-domain solver (CST Microwave Studio).
The materials were modeled with full losses: gold with $\sigma=\SI{4.1e7}{S/m}$ and high-resistivity silicon with a permittivity of $\epsilon_r=11.7$ and $\tan\delta=\SI{1.7e-5}{}$ according to \cite{Dai2004}. Fig.\,\ref{fig:load_Hfields_3freqs} shows the magnetic-field intensity for three frequencies: below, above, and well above the cut-off frequency.



The PGL load was fabricated using a silicon-on-insulator SOI) wafer, which consists of a 10-$\mu$m-thick high-resistivity (>$\SI{10}{k\ohm.cm}$) device layer, 1-$\mu$m buried oxide layer grown on a high resistivity 400-$\mu$m thick silicon handle layer. First, the front-side PGL circuit pattern was defined using electron-beam lithography followed by evaporation of a 20/350-nm-thick Ti/Au layer. Next, the high-resistivity (>$\SI{10}{k\ohm.cm}$) silicon membrane was suspended by etching the bulk silicon below the influence area of the device using a silicon-on-insulator wafer. More details of the fabrication process can be found in \cite{CabelloSanchez2022}.
By fabricating the PGLs on an electrically-thin substrate suspended in air, radiation losses drastically drop \cite{CabelloSanchez2018} since the phase velocity of the PGL mode is less than that of the substrate modes \cite{Rutledge1980}.

\begin{figure}[t]
    \centering
    \includegraphics[width=0.95\linewidth]{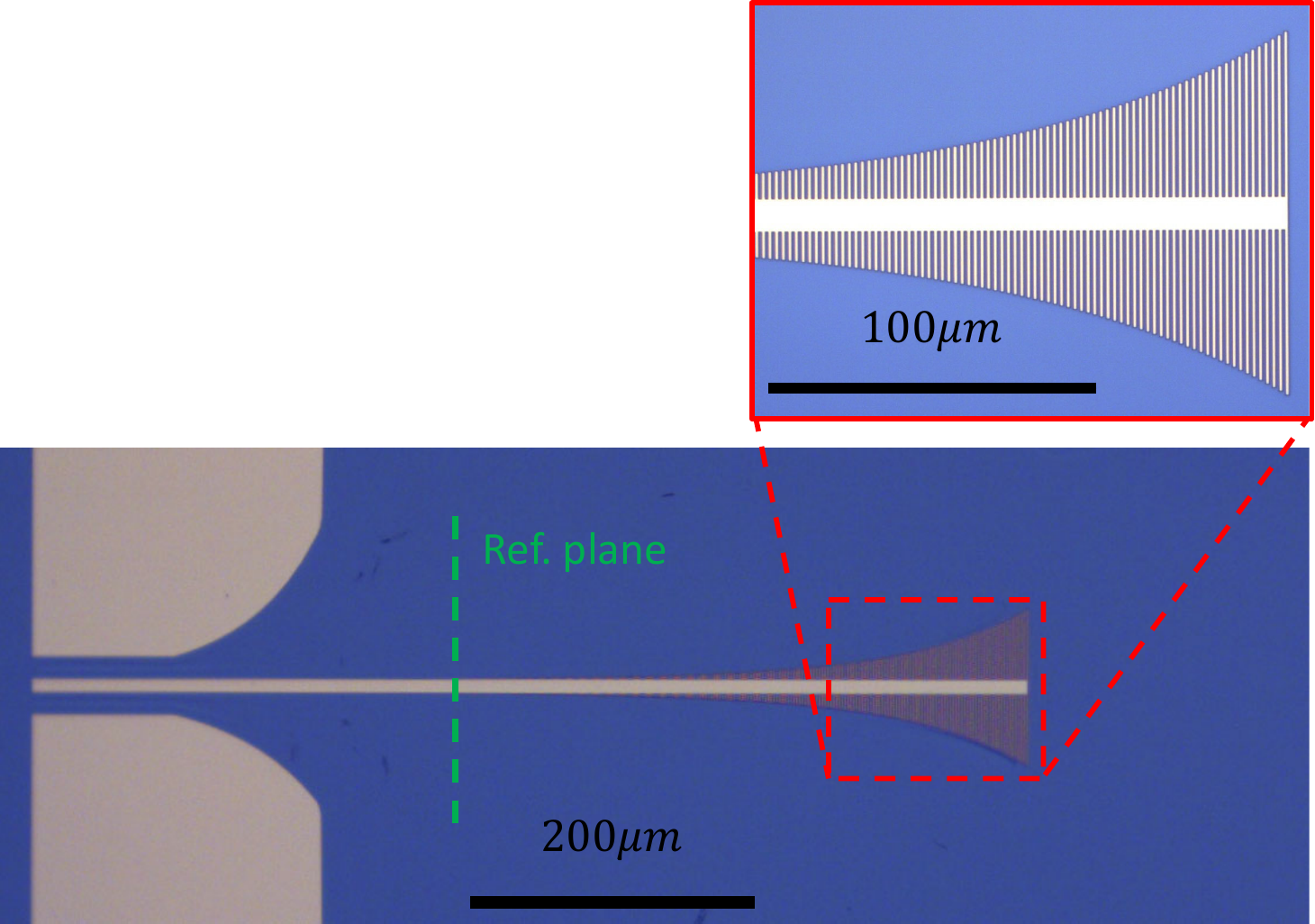}
    \caption{Micrograph of the fabricated PGL load. A CPW-PGL mode converter was used to excite the PGL using on-wafer ground-signal-ground probes. The Green dashed line shows the measurement reference plane. The inset shows a zoom-in micrograph of the load.}
    \label{fig:pgl_load2}
\end{figure}

\begin{figure}[t]
\centering
\includegraphics[width=0.95\linewidth]{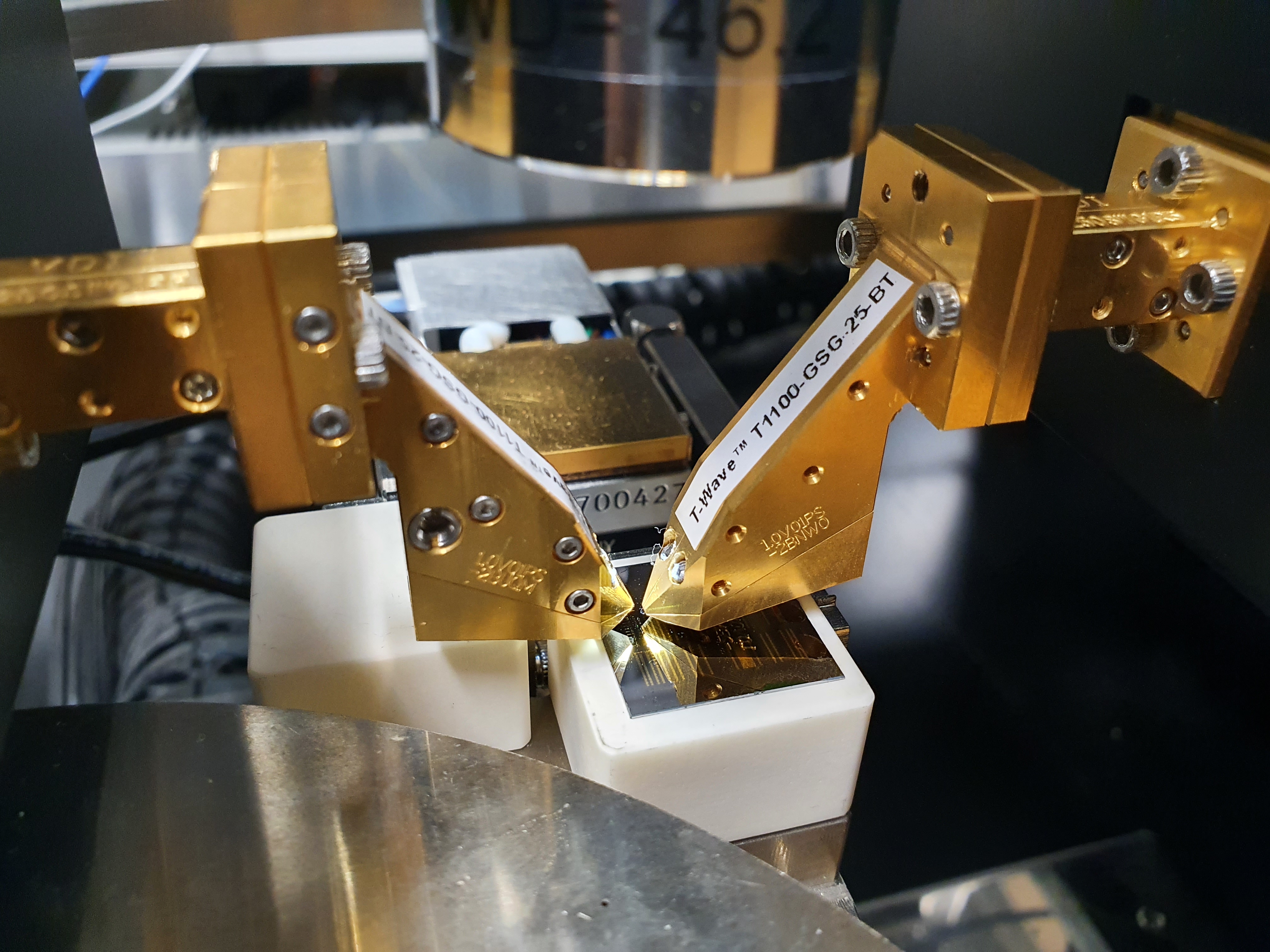}
\caption{\label{fig:setup_pic} Measurement set-up. Close-up picture of the terahertz on-wafer probes and the fabricated chip with the impedance-matched loads. Only one probe is used for the actual reflection measurements, while a two-port TRL calibration was applied.}
\end{figure}

The PGL matched load was characterized by measuring the reflection coefficient between 0.5\,THz and 1.1\,THz using a vector network analyser (Keysight N5242A) with frequency extenders (VDI WR1.5VNAX and WR1.0VNAX) and DMPI T-Wave ground-signal-ground probes \cite{Reck2011, Bauwens2014} (see Fig.\,\ref{fig:setup_pic}).
The intermediate-frequency bandwidth was set to 100\,Hz.
To excite the PGL mode using ground-signal-ground probes, we included a coplanar waveguide (CPW) mode converter, which maximizes the delivered power by minimizing losses by avoiding small strip and gap dimensions and minimizing reflections by gradually changing the characteristic impedance.
The measurements were calibrated with dedicated multi-line Thru-Reflect-Line standards for PGL \cite{Cabello-Sanchez2018} fabricated on the same chip.
Calibrating allows setting the reference plane in the PGL (see Fig.\,\ref{fig:pgl_load2}) and de-embed the coplanar waveguide transition.
The lines used for calibration have an electrical length of $\lambda_g/4$, $3\lambda_g/4$, and $11\lambda_g/4$ at the center frequency of both the WR1.5 and WR1.0 frequency bands.
Using PGL calibration standards, we characterized the matched load directly, excluding the influence of the mode converter.
The chip was placed on a ceramic chuck during measurements. After S-parameter calibration, the reflection measurement of a PGL calibration line standard indicates a noise floor of around -20\,dB as shown in Fig.~\ref{fig:S11load}. 
The suspended membrane circuit is strong enough to support the pressure from on-wafer probing.

\section{Results} \label{sec:results}

\begin{figure}[t]
\centering
\includegraphics[width=0.99\linewidth]{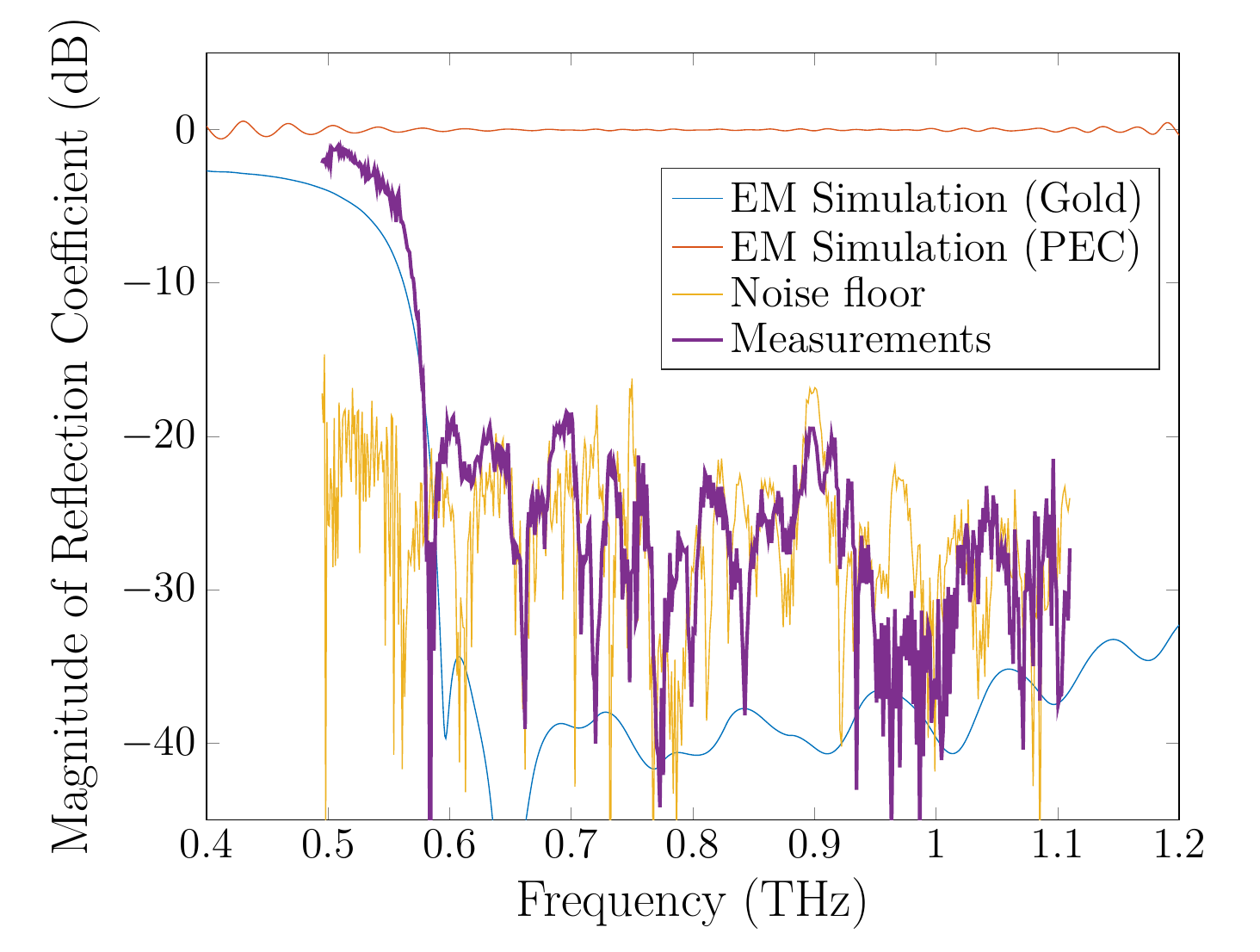}
\caption{\label{fig:S11load} Simulated and measured reflection coefficient ($S_{11}$) of the PGL matched load against frequency. The measurement of a line calibration standard is included to show the noise floor after TRL calibration}
\end{figure}

Fig.~\ref{fig:S11load} shows a comparison of the simulated and measured reflection coefficients against frequency.
There is good agreement between simulations and measurements, with a measured reflection coefficient smaller than -19\,dB between 0.58\,THz and 1.1\,THz, and around -25\,dB on average. However, the noise floor for reflection measurements in our setup limits the characterization of the load below -20\,dB.
The presented simulated results of the PGL load made out of a perfect electric conductor (PEC) show a high reflection coefficient, implying that the conductor loss and not radiation loss produce a good impedance match.
Both simulations and measurements show a cut-off in the magnitude of the reflections around \SI{0.58}{THz}. At the cut-off frequency, the length of the matched load is approximately $1.2\lambda_g$, where $\lambda_g$ is the guided wavelength (PGL), and the longest corrugation length is $0.22\lambda_c$. The electric energy stored at the tips of the corrugations can be modeled as being terminated in a capacitive susceptance, and thus the cut-off frequency may be down-shifted \cite{Collin1991Corrugated}. Simulation results of the load show a frequency bandwidth of over a decade, ultimately affected by the onset of substrate modes.

\section{Conclusion} \label{sec:concl}

In conclusion, we have demonstrated a broadband, impedance-matched load for planar Goubau lines at terahertz frequencies based on a densely and exponentially-tapered corrugated line. In the frequency range between \SI{0.5}{THz} and \SI{1.1}{THz}, the measured return loss is at least 19\,dB. This load has the advantage of having a simple design and requires no additional low-conductivity thin-film materials.
The measurements of the proposed load show high performance compared to the state-of-the-art results, shown in Table\,\ref{tab:table_comparison}.
Future work could apply a detailed parameter analysis and design methodology for the trade-off between return loss, bandwidth, and physical size of the impedance-matched load. \\

\begin{table}[htbp]
\centering
\caption{\bf State-of-the-art in PGL loads}
\begin{tabular}{lcccc}
\hline
References&\cite{xu2011}&\cite{Treizebre2012}&\cite{LeZhang2022a}&\textbf{This work}\\
\hline
Frequency (GHz) & 40--65 & 60--325 & 195--500 & 580--1100\\
Return loss (dB) & >12 & >13 & >20$^*$ & >19\\
Size & $0.16\lambda_0^2$ & $0.03\lambda_0^2$ & $0.05\lambda_0^2$ & 
$0.16\lambda_0^2$ \\
Resistive material & Ti & Ni:Cr & Ni & -- \\
\hline
\end{tabular}
\caption*{$^*$simulation results}
  \label{tab:table_comparison}
\end{table}



%

\appendices

\section*{Acknowledgment}

The authors would like to thank M. Myremark for machining parts for the measurement setup; D. Jayasankar, A. Moradikouchi, and P. Starski for their valuable feedback on the manuscript. The devices were fabricated and measured in the Nanofabrication Laboratory and Kollberg Laboratory, respectively, at the Chalmers University of Technology, Gothenburg, Sweden.

\ifCLASSOPTIONcaptionsoff
  \newpage
\fi



%


\bibliographystyle{IEEEtran} 

\bibliography{IEEEfull,bibl}

\end{document}